\documentclass[a4paper,11pt]{article}
\usepackage{jinstpub} % for details on the use of the package, please see the JINST-author-manual
\usepackage{lineno}
\usepackage{siunitx}
\usepackage{hyperref}
\usepackage{subcaption}
\usepackage{fancyhdr}

\DeclareSIUnit{\nequiv}{\ensuremath{n_\mathrm{eq}}}

\newcommand{\fluence}[1]{%
  \SI{#1}{\nequiv\per\square\centi\metre}%
}
\newcommand{\pixpitch}[2]{\ensuremath{\num{#1}\times\num{#2}\,\si{\micro\metre\squared}}}

\newcommand\blfootnote[1]{%
  \begingroup
  \renewcommand\thefootnote{}\footnote{#1}%
  \addtocounter{footnote}{-1}%
  \endgroup
}

\DeclareSIUnit{\columns}{columns}
\DeclareSIUnit{\rows}{rows}

\usepackage{tikz}
\usetikzlibrary{arrows.meta}

\title{\boldmath Recent test beam results of ATLAS ITk Pixel sensors and modules}

\collaboration[c]{on behalf of the ATLAS ITk}

\author[a,1]{A. Schmier\note{Corresponding author.},}
\author[b]{B.K. Chitroda,}
\author[c]{C. Gemme,}
\author[f]{N. Kakoty,}
\author[h]{M.N. Mantinan,}
\author[e]{H. Pang,}
\author[f]{K. Rama,}
\author[c,d]{S. Ravera,}
\author[c,d]{M. Ressegotti,}
\author[b]{A. Rummler,}
\author[f]{M. Sachdeva,}
\author[g]{M.A.A. Samy,}
\author[d]{M. Ventura,}
\author[a,i]{and J. Ye}

\affiliation[a]{INFN Trento (TIFPA),\\Via Sommarive 14, 38123, Trento, Italy}
\affiliation[b]{CERN,\\ Espl. des Particules 1, 1217, Geneva, Switzerland}
\affiliation[c]{INFN Sezione di Genova,\\Dipartimento di Fisica, Via Dodecaneso, 33, 16146, Genova, Italy}
\affiliation[d]{Università degli Studi di Genova,\\ Via Balbi, 5, 16126, Genova, Italy}
\affiliation[e]{IRFU CEA-Paris,\\Bât. 141 CEA - Saclay, 91191 Gif-sur-Yvette, France}
\affiliation[f]{IFAE Barcelona,\\Institut de Física d'Altes Energies (IFAE), Edifici Cn, Campus UAB, 08193 Bellaterra (Barcelona), Spain}
\affiliation[g]{University of Glasgow,\\Glasgow G12 8QQ, United Kingdom}
\affiliation[h]{University of Chicago,\\5801 S Ellis Ave, Chicago, IL 60637, United States}
\affiliation[i]{Università degli Studi di Trento,\\Via Sommarive 9, 38123, Trento, Italy}

\emailAdd{a.schmier@cern.ch}

\abstract{Test beam studies of ITk pixel sensors and full modules are an essential tool for assessing their performance and operational behavior. In the 2025 test beam campaigns, quad and triplet modules equipped with both pre-production and production readout chips were evaluated. For the first time, triplet modules were tested in beam conditions, both before and after irradiation. Measurements were performed at multiple incident angles to study charge-collection efficiency under conditions representative of actual detector operation. The 2025 test beam dataset contains several important new elements, enabling further progress toward the full qualification of ITk pixel modules. This work presents an overview of the ITk pixel test beam program including recent triplet module results from the 2025 campaigns.}


\arxivnumber{2605.25670} % Only if you have one

\begin{document}
\blfootnote{Copyright 2026 CERN for the benefit of the ATLAS Collaboration. CC-BY-4.0 license.}

\maketitle

\flushbottom

\section{Introduction}
\label{sec:intro}

During the next long shutdown starting in 2026, the LHC will be upgraded to provide an expected instantaneous (leveled) luminosity of \SIrange{5e34}{7.5e34}{\per\square\centi\metre\per\second}~\cite{ZurbanoFernandez:2020cco}. The extreme luminosity of the HL-LHC will require the innermost detectors to cope with 200 pile-up events per bunch crossing and a fluence of up to \fluence{2e16}~\cite{GONELLA2023167597}.

The ATLAS ITk~\cite{GONELLA2023167597} is a full-silicon tracking detector that will replace the current ATLAS inner detector to meet the new demands regarding high radiation tolerance, high granularity, low material budget and power consumption. The detectors that comprise the ITk are undergoing test beam campaigns to evaluate their performance and operational behavior in an environment representative of true detector operation. Measurements are performed both before and after irradiation in order to emulate end-of-life operation of the detector.

\begin{figure}[htbp]
\centering
\includegraphics[width=.7\textwidth]{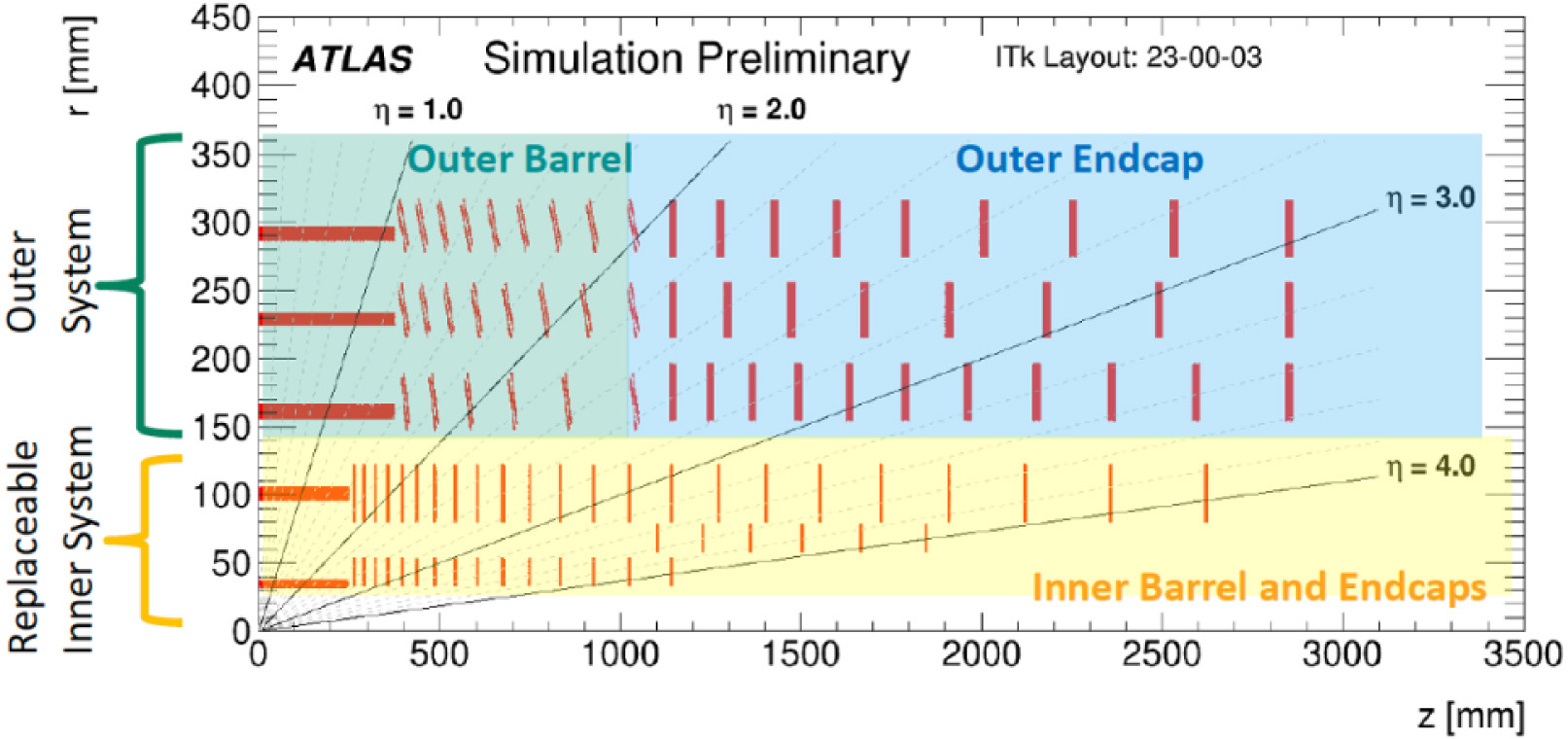}
\qquad
\caption{Schematic of the ITk pixel system showing the inner system instrumented with 3D sensor modules in yellow and the outer system instrumented with planar sensor modules in blue and green.~\cite{BUTTAR2025169978} (based on~\cite{ATL-PHYS-PUB-2021-024})\label{fig:itkinner}}
\end{figure}

This work focuses on the  ITk pixel detector~\cite{pixeltdr} which will use two different pixel sensor technologies depending on location within the detector as shown in figure~\ref{fig:itkinner}. Both technologies are realized as hybrid modules in which the sensor is bump-bonded to a readout chip. The outer layers will be instrumented with quad modules using n-in-p planar sensors. Building on the success of the current ATLAS Insertable B-Layer (IBL)~\cite{ibl}, the innermost layer and rings will be instrumented with triplet modules using 3D sensors, a technology that has demonstrated extreme radiation tolerance. 

\section{3D Triplet Modules}
\label{sec:triplets}

\begin{figure}[htbp]
    \centering
    \begin{subfigure}[t]{0.47\textwidth}
        \centering
        \vspace{0pt}
        \includegraphics[height=0.25\textheight]{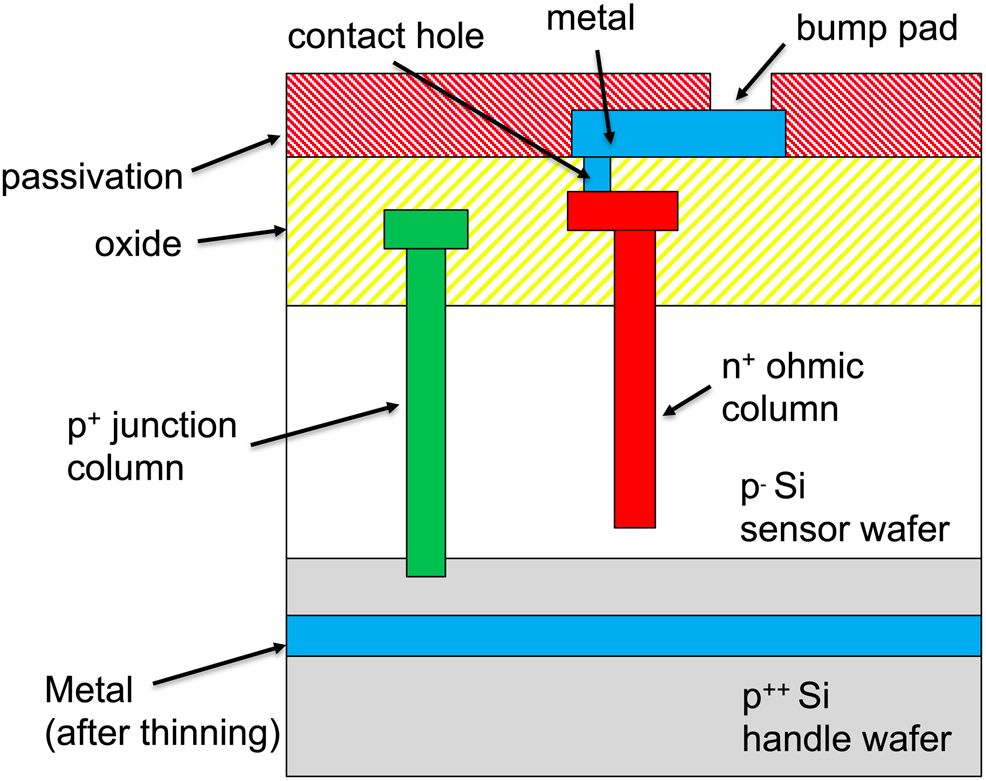} 
        \caption{} \label{fig:3da}
    \end{subfigure}
    \hfill
    \begin{subfigure}[t]{0.47\textwidth}
        \centering
        \vspace{3pt}
        \includegraphics[height=0.27\textheight]{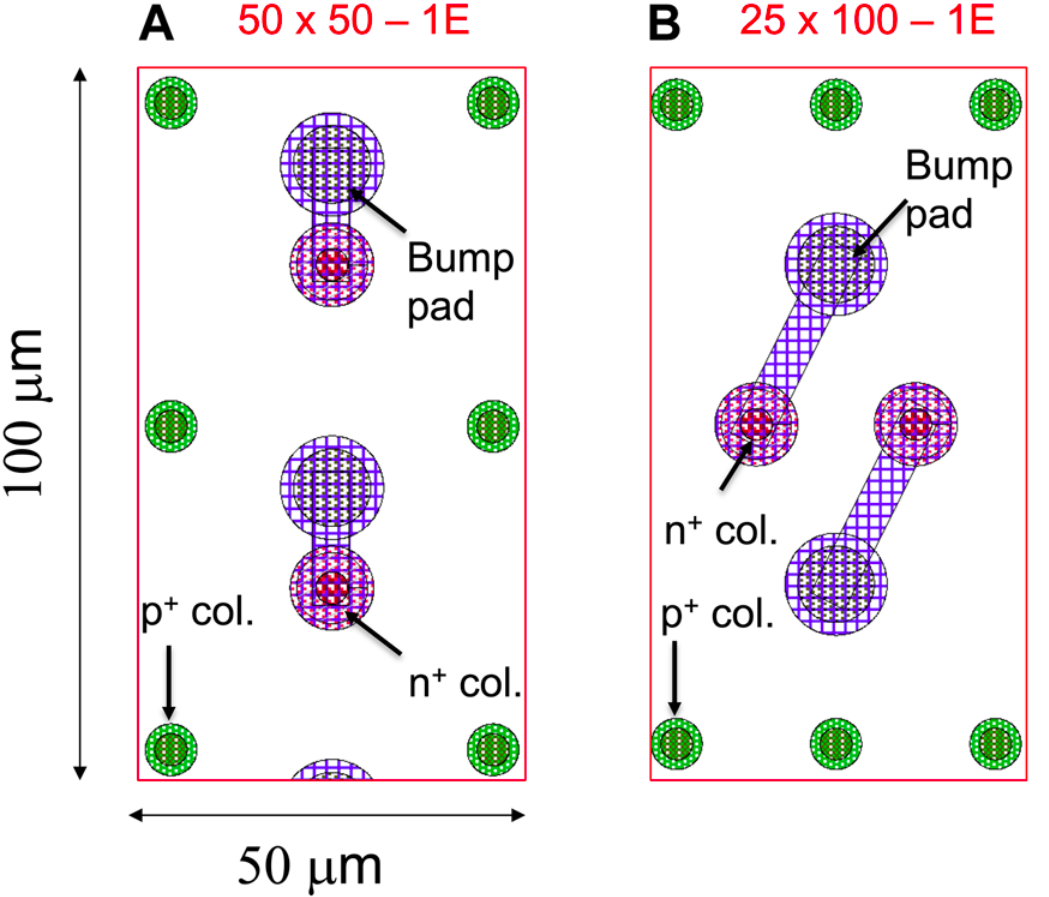} 
        \caption{} \label{fig:3db}
    \end{subfigure}
    \caption{(a) Cross section of a 3D pixel sensor. (b) Top view of both 3D sensor geometries, each showing two pixels. For the \pixpitch{50}{50} geometry, the two pixels are stacked vertically, while for the \pixpitch{25}{100} geometry, the two pixels are laterally adjacent.~\cite{3d_atlas_upgrade}\label{fig:3dsensors}}
\end{figure}

The 3D sensors developed for the ATLAS ITk use single-sided technology. As seen in figure~\ref{fig:3da}, both electrode types are etched vertically into p-type substrate from the same side and filled with polycrystalline silicon. The ohmic electrodes pass through the substrate to the handle wafer, allowing for backside biasing, while the junction electrodes stop \SI[parse-numbers=false]{\sim 25}{\micro\metre} from the handle wafer. The inter-electrode spacing is no longer dependent on the sensor thickness and can be made very small, resulting in high radiation hardness. Figure~\ref{fig:3db} shows the two different pixel geometries used in the ITk in order to adapt to specific geometric needs based on position in the detector. Pixels with a \pixpitch{25}{100} layout are used to construct linear triplets for the innermost barrel layer, while pixels with a \pixpitch{50}{50} layout are used to construct ring triplets. At the sensor-level, the pixel matrix is terminated with p-type column fences around the perimeter. At the pixel-level, electrodes are isolated from each other using p-spray~\cite{pspray}.

\begin{figure}[htbp]
\centering
\includegraphics[width=.5\textwidth]{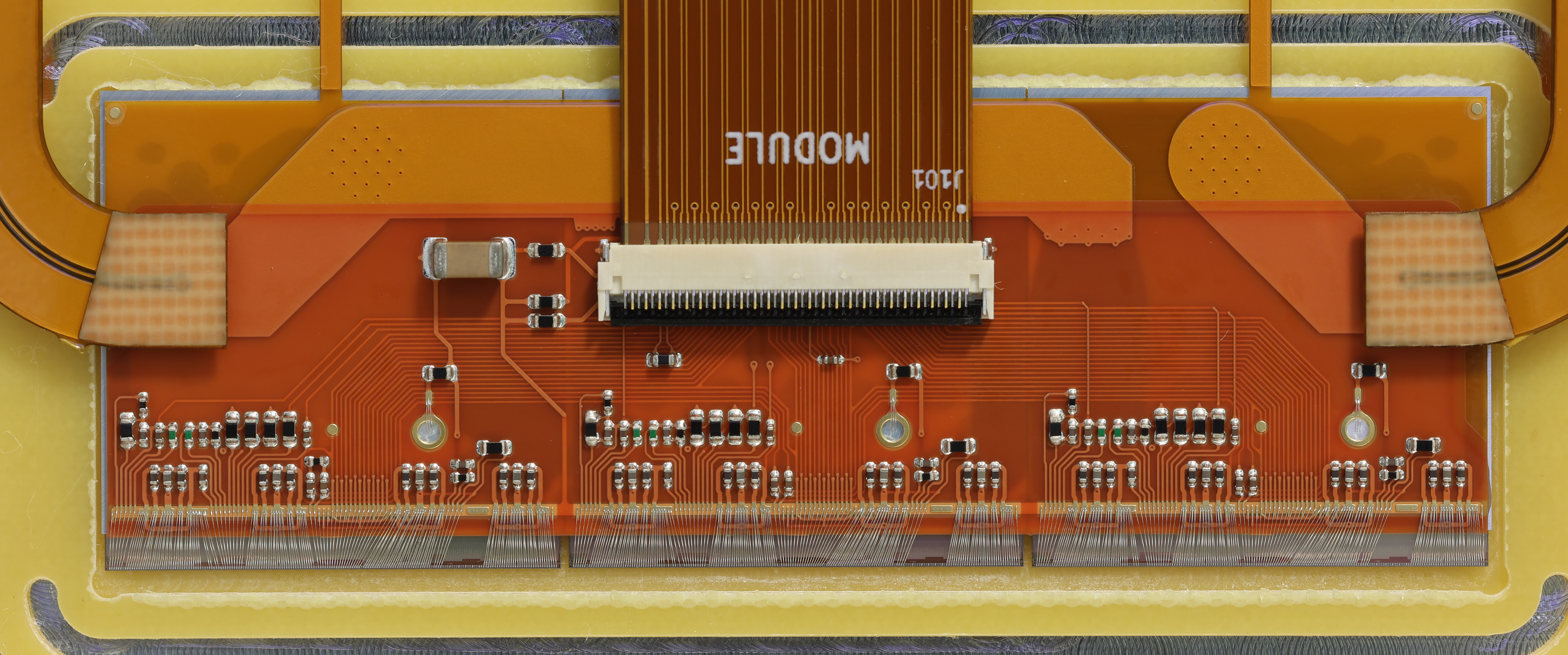}
\qquad
\caption{Fully assembled linear triplet module.\label{fig:modulephoto}}
\end{figure}

All pixel sensor types are bump bonded to RD53 readout ASICs~\cite{Alimonti_2025} utilizing a differential front-end. Pre-production modules use the RD53B chip, i.e. ITkPixV1.1, while production modules use the RD53C, i.e. ITkPixV2. Each chip has a matrix of 400~columns $\times$ 384~rows, resulting in \SI[group-separator = {,}]{153600}{} pixels per chip. The electrode pitch is fixed to \pixpitch{50}{50}, requiring some surface-level readout electrode routing at the pixel to adapt to the \pixpitch{25}{100} geometry, shown in blue in figure~\ref{fig:3dsensors}. The final product, shown in figure~\ref{fig:modulephoto}, combines three hybrid sensors to form a triplet module. The hybrid sensors are glued and wire bonded to a flexible PCB which provides power and communication via pigtail adapters.

\section{Test Beam Setup}
\label{sec:telescope}

\begin{figure}[htbp]
    \centering
    \begin{subfigure}[t]{0.56\textwidth}
        \centering
        \begin{tikzpicture}
            \node[anchor=south west, inner sep=0] (img) at (0,0)
    {\includegraphics[width=\linewidth]{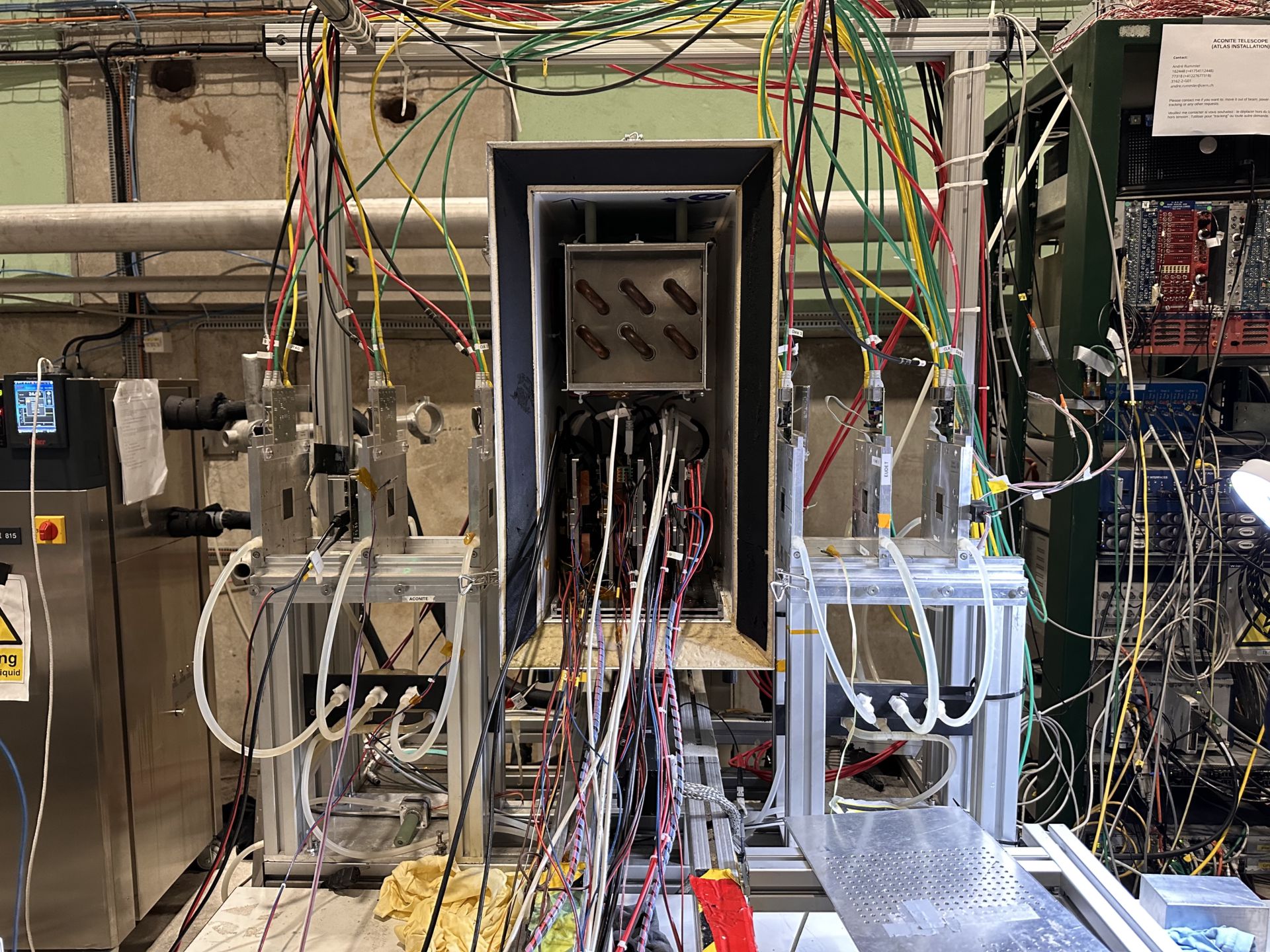}};

  % normalised coordinates:
  % (0,0) = bottom left of image, (1,1) = top right
  \begin{scope}[x={(img.south east)}, y={(img.north west)}]

    % vertical position of the beam, chosen to pass through the black squares
    \def\beamy{0.47}

    \tikzset{
      beam/.style={
        draw=cyan!70!blue,
        line width=1.8pt,
        line cap=rectangular,
        %preaction={draw=white, line width=5.2pt, opacity=0.85}
      }
    }

    % label
    \node[
      fill=white,
      fill opacity=0.85,
      text opacity=1,
      rounded corners=1pt,
      inner xsep=2pt,
      inner ysep=1pt,
      font=\scriptsize
    ] at (0.08,\beamy+0.035) {beam};

    % interrupted beam line:
    % left of outer-left plane
    \draw[beam] (-0.05,\beamy) -- (0.197,\beamy);

    % between the two left telescope planes
    \draw[beam] (0.23,\beamy) -- (0.283,\beamy);

    % between left telescope arm and DUT box
    \draw[beam] (0.31,\beamy) -- (0.37,\beamy);

    % between DUT box and right telescope arm
    \draw[beam] (0.6325,\beamy) -- (0.679,\beamy);

    % between the two right telescope planes
    \draw[beam] (0.702,\beamy) -- (0.74,\beamy);

    % right of outer-right plane, with arrowhead
    \draw[beam, -{Latex[length=4mm,width=2.6mm]}]
      (0.8,\beamy) -- (1.07,\beamy);

  \end{scope}
\end{tikzpicture}        
        \caption{} \label{fig:telescopea}
    \end{subfigure}
    \hfill
    \begin{subfigure}[t]{0.36\textwidth}
        \centering
        \includegraphics[width=\linewidth, origin=c]{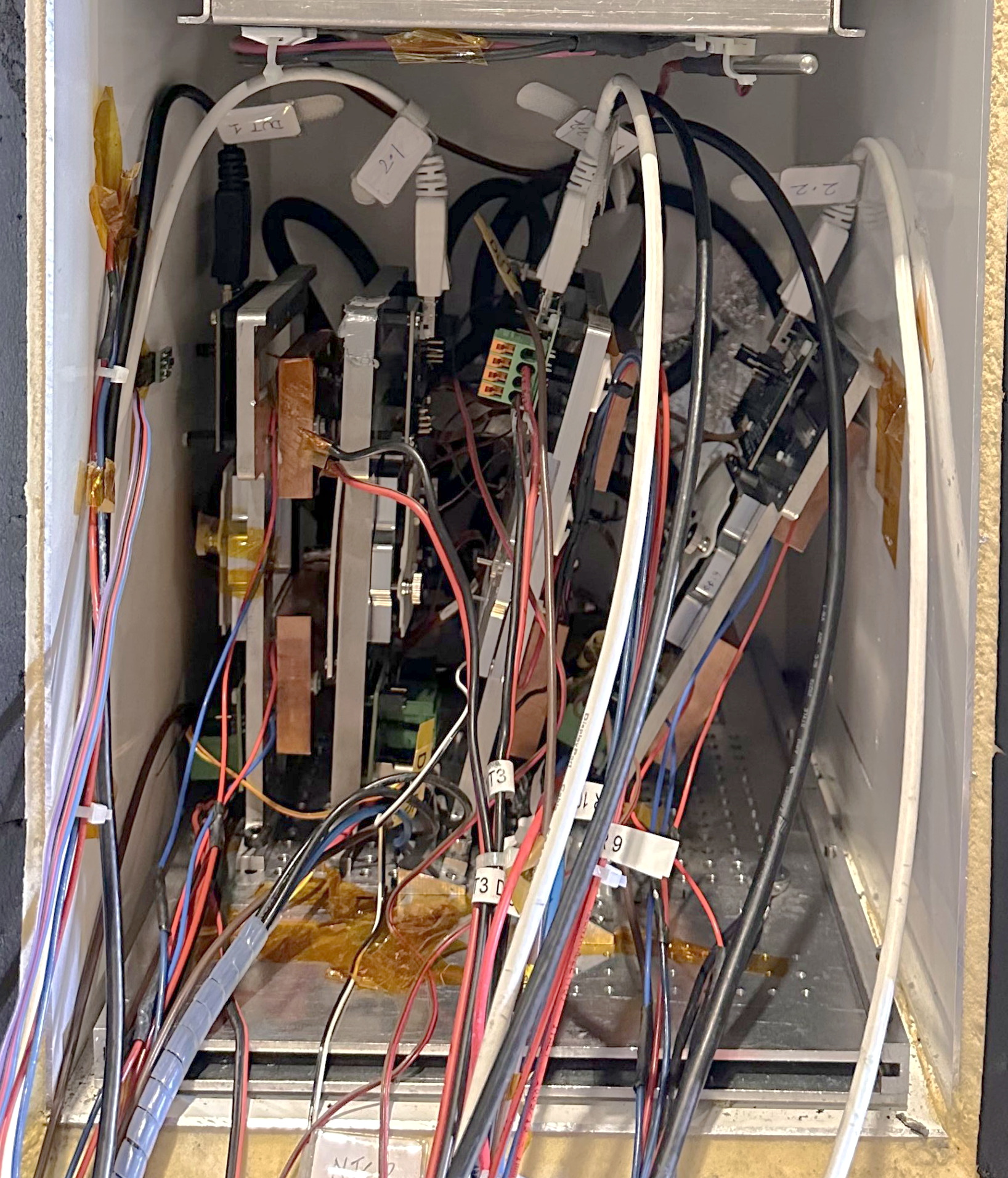}
        \caption{} \label{fig:telescopeb}
    \end{subfigure}
    \caption{(a) ACONITE beam telescope at the CERN SPS H6 beamline. (b) Four modules arranged inside the climate-controlled box with all connections in place. The first two modules from the left are perpendicular to the beam. The third is tilted \SI{15}{\degree} and the fourth \SI{30}{\degree} relative to the beam.\label{fig:telescopesetup}}
\end{figure}
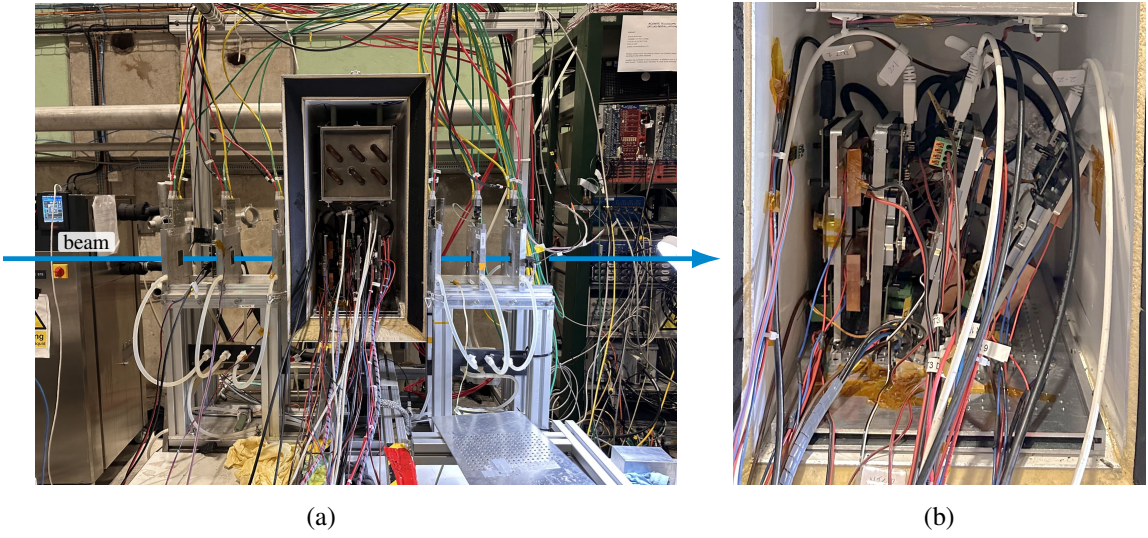

Recent test beam campaigns for the ATLAS ITk were carried out using the EUDET type ACONITE telescope~\cite{Jansen2016EUDET} at the CERN SPS H6 beamline~\cite{Banerjee:NorthArea:2021} using a \SI{120}{\giga\electronvolt} pion beam. The ACONITE telescope, shown in figure~\ref{fig:telescopea}, consists of six MIMOSA26~\cite{HUGUO2010480} tracking planes, a timing reference, and a trigger logic unit (TLU)~\cite{Cussans:EUDET-MEMO-2009-04}. The MIMOSA26 planes demonstrate an excellent tracking resolution of \SI{\sim 5.3}{\micro\metre} due to their pixel size of \pixpitch{18.4}{18.4} but their timing is significantly worse as they implement a rolling shutter system with an integration time of \SI{\sim 230}{\micro\second} for a double frame. A separate timing plane is used in order to select the fraction of tracks recorded by the MIMOSA26 planes which are in-time for the devices under test (DUTs) according to the trigger scheme. This timing plane is a single chip card mounted on one of the telescope planes, consisting of a 3D sensor with \pixpitch{50}{50} geometry bonded to an ITkPixV2 front-end chip. The trigger is generated by the TLU using the coincidence of up to four scintillators and distributed to all the devices. EUDAQ2~\cite{Liu_2019} is used to orchestrate between the separate DAQ software stacks, to collect the data from telescope, TLU and DUTs. The YARR\cite{heim_2025_15007379} DAQ software (in form of a EUDAQ2 producer) is used together with SPEC PCIe cards to read out the DUTs as well as the timing plane. 

The DUTs are placed in a climate-controlled box between two sets of three telescope planes. Four or more modules can be mounted in the box at different angles using custom-made wedges or a rotating block to mimic final installation conditions, resulting in incident angles ranging from \SIrange{0}{\sim 30}{\degree} (\SI{82}{\degree}). Cooling is realized by a two-stage cooling system, one for the ambient temperature within the box, and the other for direct cooling. In addition, the box is continually flushed with pre-cooled nitrogen, allowing the modules to reach temperatures below \SI{-25}{\degree}C. The test beam setup provides the ability to remotely monitor all environmental conditions and NTC temperatures of the modules.

\section{Reconstruction and Analysis}
\label{sec:reco}

The data for this campaign are reconstructed using version 2.0 of the Corryvreckan~\cite{Dannheim_2021} software package. At the time of this work, the analysis and calculation of systematic uncertainties contributing to these test beam data are still underway. Thus, exact numbers will not be discussed here. Instead, a general overview of the reconstruction process is outlined.

The first analysis step is the masking of noisy and dead pixels. Any pixel that registers significantly more or fewer hits than the average pixel is masked. The limits are set with a large margin to avoid cutting on the non-uniform beam distribution. Neighboring hit pixels are clustered together. Correlations are found between the reference plane and all the other planes, and a rough alignment of both telescope and DUT planes is performed using the correlated cluster information. Following this rough alignment, a more stringent alignment is performed, first for the telescope planes, then for the DUTs. Clusters are associated with tracks constructed using hit information from the telescope planes. After the telescope planes are aligned, the same procedure is performed for the DUTs.

\section{Results}
\label{sec:results}

The ITk is dedicated to precision measurements of particle position for track reconstruction. As such, one of the most important measurements to be performed is the total hit efficiency, referred to simply as "efficiency" for the remainder of this work. The hit efficiency is defined as the fraction of selected telescope tracks with an associated DUT cluster within the fiducial region and matching window. The requirement posed by the ITk for irradiated 3D sensors is a minimum efficiency of \SI{96}{\percent} at perpendicular incidence.  In Corryvreckan, the efficiency is measured by comparing cluster positions in the DUT with the interpolated track position.

\subsection{Effects of Radiation}
\label{sec:radeffects}

Due to the high fluence that triplet modules must withstand over their lifetime, test beam studies with irradiated modules are essential. Modules for the 2025 test beam campaigns were irradiated at the IRRAD proton facility at the CERN PS with a \SI{24}{\giga\electronvolt} proton beam and at the RARiS facility at Tohoku University in Japan with proton beams up to \SI{80}{\mega\electronvolt}. In this work, the FBK+LND \pixpitch{25}{100} module was irradiated at the RARiS facility and received a uniform fluence of \fluence{1.1e16}. The FBK+IZM \pixpitch{50}{50} module was irradiated at IRRAD with a beam of \SI{6}{\milli\metre} FWHM. To compensate for the small beam diameter, the module was tilted by \SI{\sim 50}{\degree} about its vertical axis and horizontally scanned to reach uniform irradiation on the horizontal axis. The resulting average fluence was approximately \fluence{0.4e16} with a peak of approximately \fluence{1.0e16}. During test beam, the module was positioned such that the beam spot hit the module at the point of highest fluence. This is reflected in the plots below where the fluence is reported as \fluence{1.0e16} MAX for the peak fluence. The fluence was evaluated by placing aluminum foil under the sensor during irradiation. Nuclear spallation of aluminum results in the creation of gamma-emitting nuclides, which can then be measured with gamma spectrometry.

\begin{figure}[htbp]
\centering
\includegraphics[width=.7\textwidth]{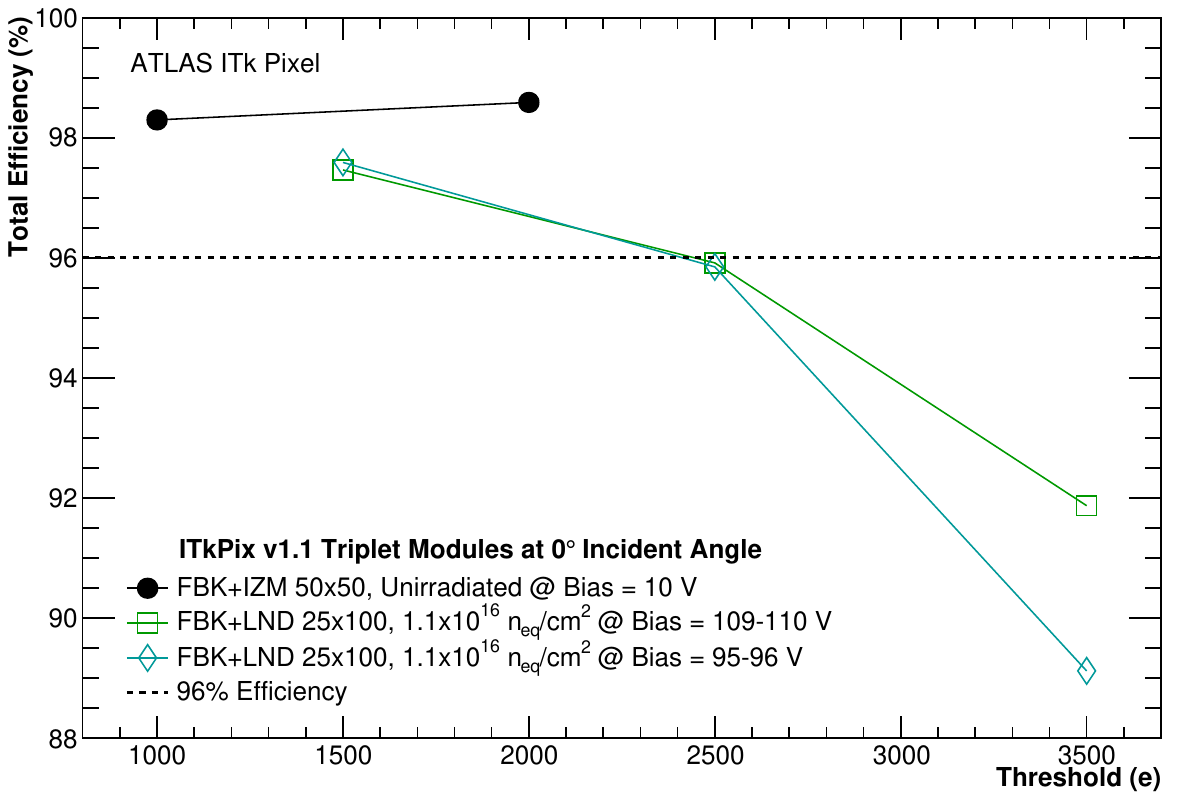}
\qquad
\caption{Efficiency versus threshold for an unirradiated module and a module irradiated to \fluence{1.1e16}\label{fig:effvsthr} at two different bias voltages.}
\end{figure}

\begin{figure}[htbp]
    \centering
    \begin{subfigure}[t]{0.45\textwidth}
        \centering
        \includegraphics[width=\linewidth]{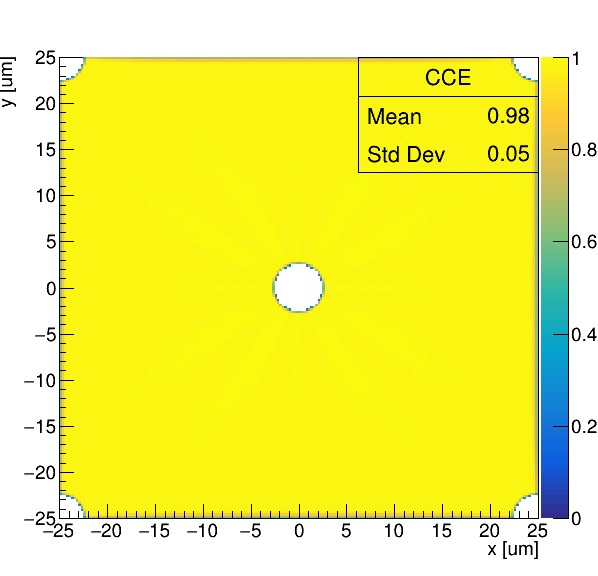} 
        \caption{} \label{fig:ccea}
    \end{subfigure}
    \hfill
    \begin{subfigure}[t]{0.45\textwidth}
        \centering
        \includegraphics[width=\linewidth]{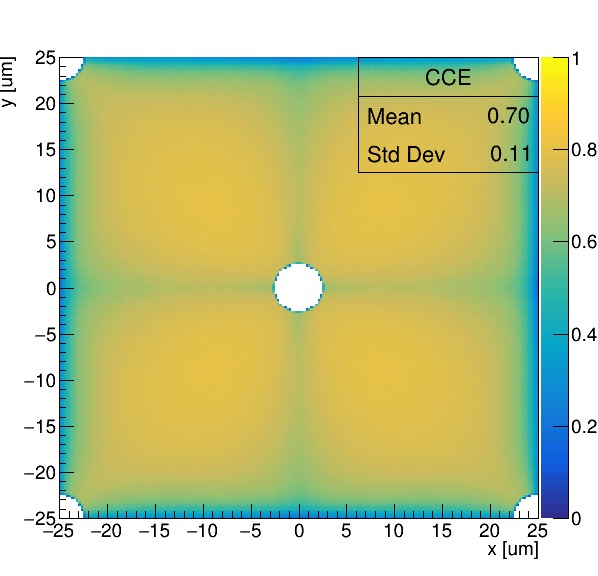} 
        \caption{} \label{fig:cceb}
    \end{subfigure}
    \caption{(a) Top-view Allpix\textsuperscript{2}~\cite{allpix2} simulation of charge collection efficiency in a single pixel for an unirradiated \pixpitch{50}{50} 3D pixel. (b) Pixel irradiated to \fluence{1e16} using the LHCb Radiation Damage Model~\cite{lhcbrad}. The electric field and weighting field are calculated with TCAD~\cite{tcad} using exact sensor geometries. The trapping effect is modeled with effective trapping time~\cite{trapping}. Electrode columns cause the efficiency drop to zero in  the center and at the edges. For both simulations, the bias voltage is \SI{100}{\volt}.\label{fig:cce}}
\end{figure}

The pixel threshold, specified as the number of deposited electrons, is the minimum amount of charge deposited in a hit pixel required for a signal to be registered as a hit. Figure~\ref{fig:effvsthr} shows that before irradiation, changing the threshold within the studied range has a negligible effect on the efficiency. After irradiation, increasing the threshold progressively decreases the efficiency. This can be attributed to decreased charge collection caused by radiation-induced trapping in the silicon~\cite{trapping}. Radiation damage also changes the effective space charge, increasing the bias voltage needed for full depletion of the active volume. Figure~\ref{fig:cce} provides a simulation of the charge collection efficiency (CCE) before and after irradiation at \fluence{1e16}, where the irradiated sensor in figure~\ref{fig:cceb} shows notably reduced CCE. Adjusting the threshold is a balancing act that must be studied under real-world conditions. Lower thresholds allow for a high chance of seeing a hit in an irradiated pixel. However, irradiation increases leakage current, therefore increasing noise and the possibility of registering fake hits.

\subsection{Mitigating Radiation Damage: Bias Voltage}
\label{sec:bias}

\begin{figure}[htbp]
\centering
\includegraphics[width=.7\textwidth]{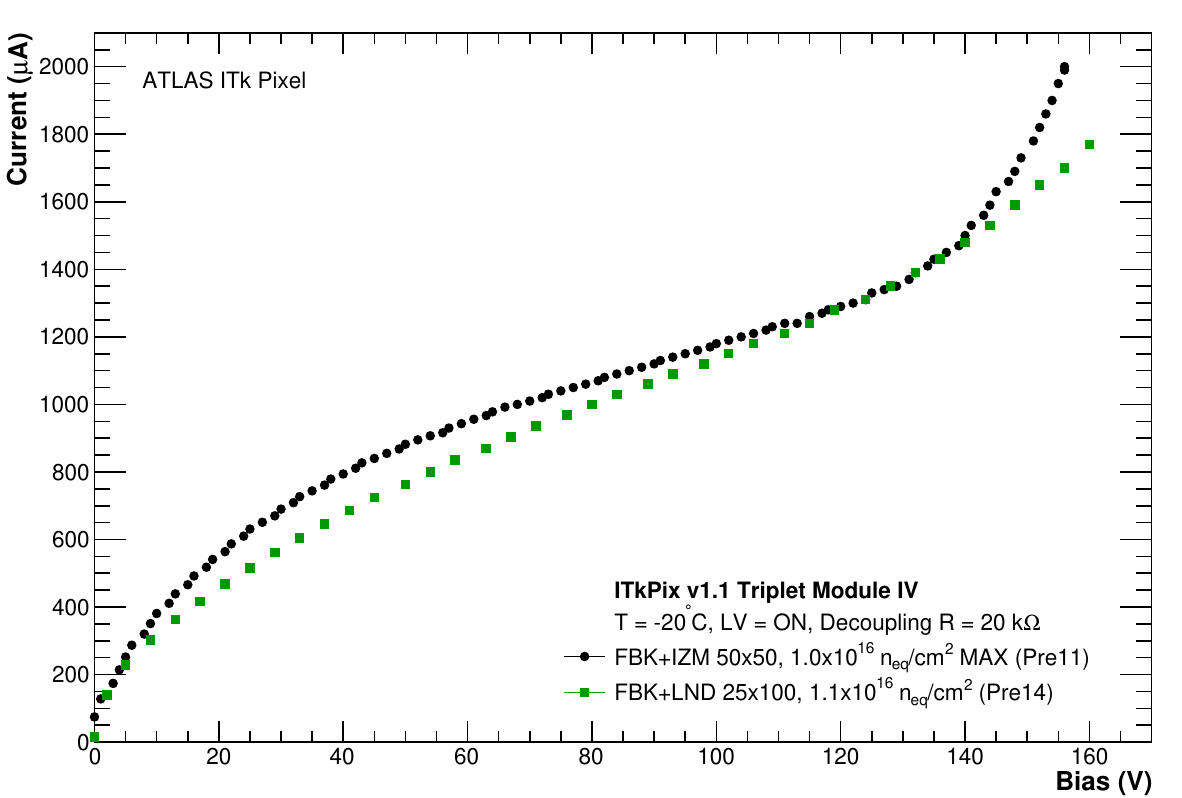}
\qquad
\caption{Leakage current versus bias voltage for two irradiated pre-production modules collected during the 2025 test beam campaigns.\label{fig:leakagevbias}}
\end{figure}

\begin{figure}[htbp]
\centering
\includegraphics[width=.7\textwidth]{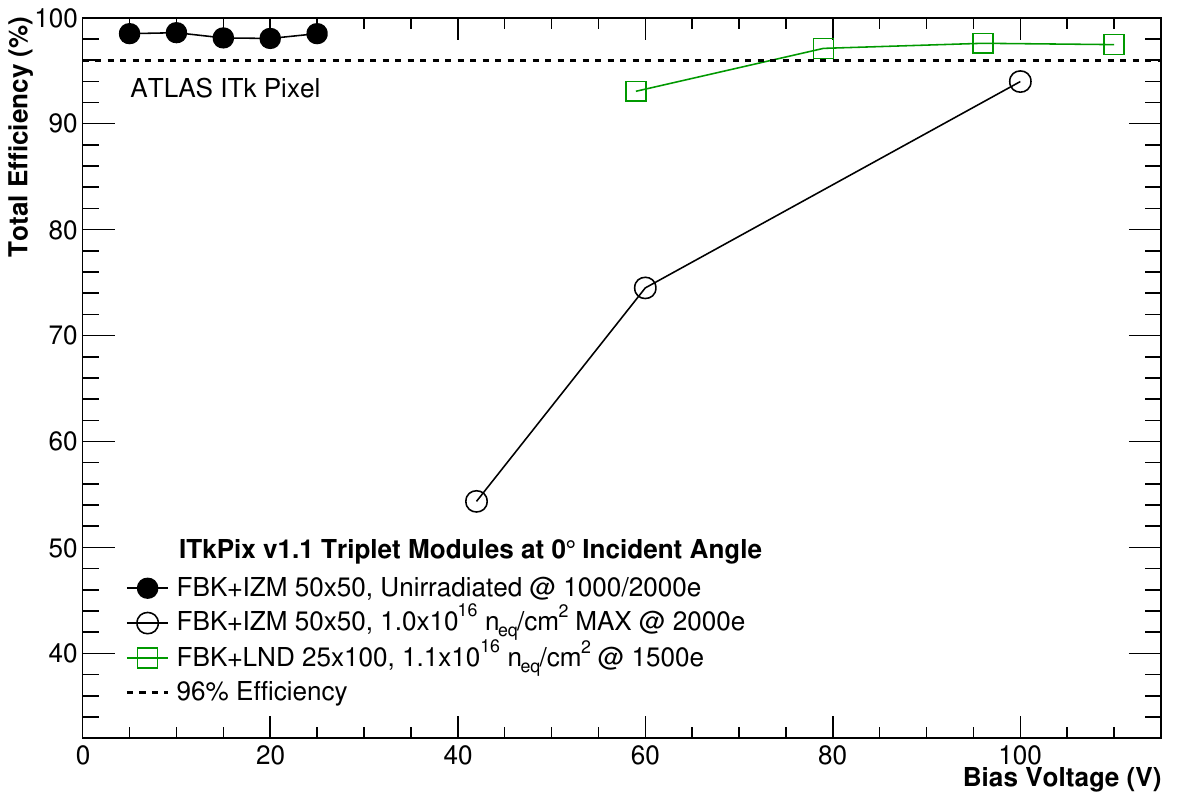}
\qquad
\caption{Efficiency versus bias voltage for an unirradiated sensor and two irradiated sensors. One sensor has \pixpitch{50}{50} pixels and was irradiated non-uniformly to an average fluence of \fluence{0.4e16} and a maximum fluence of \fluence{1e16}. The other sensor has \pixpitch{25}{100} pixels and was irradiated to \fluence{1.1e16}.\label{fig:effvbias}}
\end{figure}

Increased bias voltage can be used in order to reach full depletion again. The behavior of two modules studied during the 2025 test beam campaigns is shown in figure~\ref{fig:leakagevbias}. The point where the sensor leakage current begins to follow an exponential trend with respect to the bias voltage is referred to as breakdown. For these modules, this occurs at a bias voltage of around \SIrange{140}{150}{\volt}. In order to avoid damage, care must be taken not to push the sensors into breakdown. However, voltages up to \SI{135}{\volt} have been studied in test beam without reaching breakdown. Figure~\ref{fig:effvbias} shows the performance of triplet modules with respect to bias voltage. Before irradiation, an efficiency of over \SI{98}{\percent} is achieved with very low bias voltage. After radiation damage, an efficiency of \SI{94}{\percent} is already reached with a bias voltage of \SI{100}{\volt}.

\subsection{Mitigating Radiation Damage: Incident Angle}
\label{sec:angle}

Another way to increase efficiency after irradiation is by changing the angle of the modules with respect to the beam. Figure~\ref{fig:telescopeb} shows an example of modules positioned in the telescope at varying angles. Under the correct conditions, the path length of the particle within the active region is increased, depositing more charge in a single pixel. If instead the particle hits the pixel close to its edge or if the angle is large enough, the track will be split into two or more pixels, increasing the number of hit pixels, but depositing less charge per pixel.

\begin{figure}[htbp]
\centering
\includegraphics[width=.7\textwidth]{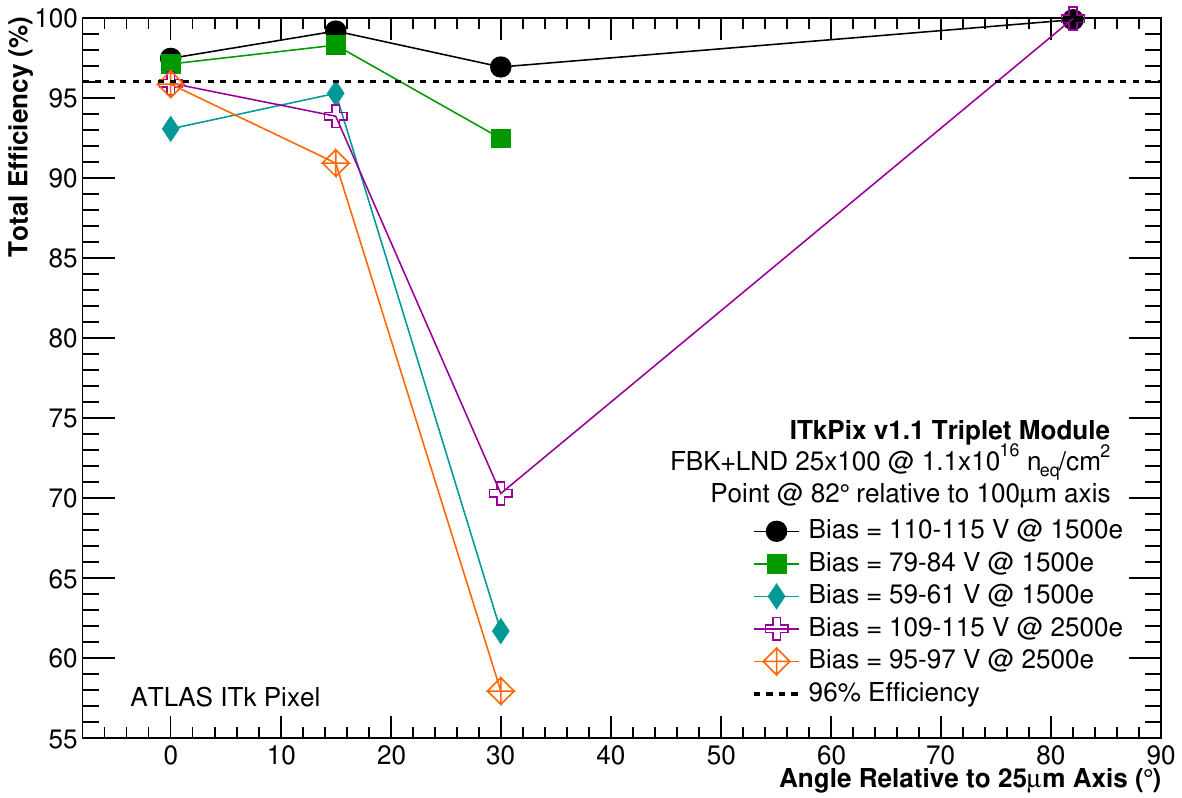}
\qquad
\caption{Efficiency versus angle for an irradiated triplet for varying bias voltages and thresholds. The final point at \SI{82}{\degree} has the module tilted so that the particle travels along the \SI{100}{\micro\metre} edge, while the remaining points have the particle traveling along the short edge. Bias voltages varied by a few volts across angle and threshold configurations.~\label{fig:effvangle}}
\end{figure}

Figure~\ref{fig:effvangle} shows the efficiency as a function of incident angle along the \SI{25}{\micro\metre} axis of an irradiated module with \pixpitch{25}{100} pixel geometry. At lower thresholds, increasing the angle from \SIrange{0}{15}{\degree} increases the efficiency. However, increasing from \SIrange{15}{30}{\degree} instead reduces the efficiency due to decreased in-pixel path length. At higher thresholds, an increase of even \SI{15}{\degree} results in a reduced efficiency since any charge sharing brings the per-pixel deposit closer to the threshold, reducing the hit probability.

Shallow incidence configurations, where particles reach the sensor at near-grazing angles, are representative of the conditions seen by the innermost barrel layer (L0), the endcap modules at large pseudorapidity, and the Pixel Luminosity Rings (PLR). The \SI{30}{\degree} scans already showed a marked efficiency drop after irradiation, motivating a dedicated point at \SI{82}{\degree}, where the particle travels along the \SI{100}{\micro\metre} pixel axis instead of the \SI{25}{\micro\metre} one. In that configuration the efficiency exceeds \SI{99}{\percent} for both studied thresholds since many pixels will experience the maximum possible path length. 

\section{Discussion}
\label{sec:discussion}

Test beam studies have been performed in order to investigate the performance of the new ATLAS ITk under conditions representative of real detector operations. Recent test beam studies performed at CERN include important measurements of 3D triplet modules. The effects of radiation on these modules are apparent in the total hit efficiency. Two methods have been studied in test beam to mitigate the effects of radiation. Increasing the bias voltage allows sensors to reach full depletion after irradiation, resulting in a high efficiency well before the breakdown voltage. Tilting the modules at an angle with respect to the beam mimics real detector conditions, particularly in the rings. At low thresholds, small angles improve the efficiency of irradiated modules. At high threshold values or large angles, the efficiency drops with respect to those at normal incidence.

% Bibliography

%% [A] Recommended: using JHEP.bst file
\bibliographystyle{JHEP}
\bibliography{biblio.bib}

@ARTICLE{3d_atlas_upgrade,
    AUTHOR={Terzo, S.  and others},
    TITLE={Novel {3D} {P}ixel {S}ensors for the {U}pgrade of the {ATLAS} Inner Tracker},
    JOURNAL={Frontiers in Physics}, 
    VOLUME={Volume 9 - 2021},
    YEAR={2021},
    URL={https://www.frontiersin.org/journals/physics/articles/10.3389/fphy.2021.624668},
    DOI="10.3389/fphy.2021.624668",
    ISSN={2296-424X}
}

@ARTICLE{pspray,
  author={Piemonte, C.},
  journal={IEEE Transactions on Nuclear Science}, 
  title={Device simulations of isolation techniques for silicon microstrip detectors made on p-type substrates}, 
  year={2006},
  volume={53},
  number={3},
  pages={1694-1705},
  doi={10.1109/TNS.2006.872500}
}

@article{GONELLA2023167597,
title = {The {ATLAS} {ITk} detector system for the {Phase-II} {LHC} upgrade},
journal = {Nuclear Instruments and Methods in Physics Research Section A: Accelerators, Spectrometers, Detectors and Associated Equipment},
volume = {1045},
pages = {167597},
year = {2023},
issn = {0168-9002},
doi = "10.1016/j.nima.2022.167597",
url = {https://www.sciencedirect.com/science/article/pii/S0168900222008890},
author = {{L. Gonnella for the ATLAS ITk Collaboration}}
}

@article{allpix2,
title = {Allpix\textsuperscript{2}: {A} modular simulation framework for silicon detectors},
journal = {Nuclear Instruments and Methods in Physics Research Section A: Accelerators, Spectrometers, Detectors and Associated Equipment},
volume = {901},
pages = {164-172},
year = {2018},
issn = {0168-9002},
doi = "10.1016/j.nima.2018.06.020",
url = {https://www.sciencedirect.com/science/article/pii/S0168900218307411},
author = {S. Spannagel and K. Wolters and D. Hynds and N. {Alipour Tehrani} and M. Benoit and D. Dannheim and N. Gauvin and A. Nürnberg and P. Schütze and M. Vicente},
}

@article{trapping,
doi = "10.1088/1748-0221/15/11/P11018",
url = {https://doi.org/10.1088/1748-0221/15/11/P11018},
year = {2020},
month = {nov},
publisher = {},
volume = {15},
number = {11},
pages = {P11018},
author = {Mandić, I. and Cindro, V. and Gorišek, A. and Hiti, B. and Kramberger, G. and Mikuž, M. and Zavrtanik, M. and Skomina, P. and Hidalgo, S. and Pellegrini, G.},
title = {Measurements with silicon detectors at extreme neutron fluences},
journal = {Journal of Instrumentation}
}

@misc{tcad,
  author       = {{Synopsys, Inc.}},
  title        = {Technology Computer Aided Design ({TCAD})},
  url          = {https://www.synopsys.com/manufacturing/tcad.html},
  note         = {Accessed: 2026-05-20}
}

@article{Dannheim_2021,
doi = "10.1088/1748-0221/16/03/P03008",
url = {https://doi.org/10.1088/1748-0221/16/03/P03008},
year = {2021},
month = {mar},
publisher = {IOP Publishing},
volume = {16},
number = {03},
pages = {P03008},
author = {Dannheim, D. and Dort, K. and Huth, L. and Hynds, D. and Kremastiotis, I. and Kröger, J. and Munker, M. and Pitters, F. and Schütze, P. and Spannagel, S. and Vanat, T. and Williams, M.},
title = {Corryvreckan: a modular {4D} track reconstruction and analysis software for test beam data},
journal = {Journal of Instrumentation}
}

@techreport{Cussans:EUDET-MEMO-2009-04,
  author      = {Cussans, D.},
  title       = {Description of the {JRA1} Trigger Logic Unit ({TLU}), v0.2c},
  institution = {EUDET},
  type        = {EUDET Memo},
  number      = {EUDET-Memo-2009-04},
  year        = 2009,
  date        = {2009-09-11},
  url         = {https://www.eudet.org/e26/e28/e42441/e57298/EUDET-MEMO-2009-04.pdf}
}

@techreport{Banerjee:NorthArea:2021,
  author      = {Banerjee, Dipanwita and Bernhard, Johannes and Brugger, Markus
                 and Charitonidis, Nikolaos and Doble, Niels and Gatignon, Lau
                 and Gerbershagen, Alexander},
  title       = {The North Experimental Area at the {CERN} Super Proton Synchrotron},
  institution = {CERN},
  type        = {CERN Accelerator Note},
  number      = {CERN-ACC-NOTE-2021-0015},
  year        = {2021},
  month       = jul,
  doi         = "10.17181/CERN.GP3K.0S1Y",
  url         = {https://cds.cern.ch/record/2774716}
}

@article{Jansen2016EUDET,
  author  = {Jansen, Hendrik and Spannagel, Simon and Behr, J{\"o}rg and
             Bulgheroni, Antonio and Claus, Gilles and Corrin, Emlyn and
             Cussans, David and Dreyling-Eschweiler, Jan and Eckstein, Doris and
             Eichhorn, Thomas and Goffe, Mathieu and Gregor, Ingrid Maria and
             Haas, Daniel and Muhl, Carsten and Perrey, Hanno and Peschke, Richard and
             Roloff, Philipp and Rubinskiy, Igor and Winter, Marc},
  title   = {Performance of the {EUDET}-type beam telescopes},
  journal = {EPJ Techniques and Instrumentation},
  volume  = {3},
  number  = {1},
  pages   = {7},
  year    = {2016},
  doi     = "10.1140/epjti/s40485-016-0033-2",
  url     = {https://doi.org/10.1140/epjti/s40485-016-0033-2}
}

@article{lhcbrad,
title = {Development of a silicon bulk radiation damage model for {Sentaurus} {TCAD}},
journal = {Nuclear Instruments and Methods in Physics Research Section A: Accelerators, Spectrometers, Detectors and Associated Equipment},
volume = {874},
pages = {94-102},
year = {2017},
issn = {0168-9002},
doi = "10.1016/j.nima.2017.08.042",
url = {https://www.sciencedirect.com/science/article/pii/S0168900217309282},
author = {{\AA}. Folkestad and K. Akiba and M. {van Beuzekom} and E. Buchanan and P. Collins and E. Dall’Occo and A. {Di Canto} and T. Evans and V. {Franco Lima} and J. {García Pardiñas} and H. Schindler and M. Vicente and M. {Vieites Diaz} and M. Williams}
}

@article{Liu_2019,
doi = "10.1088/1748-0221/14/10/P10033",
url = {https://doi.org/10.1088/1748-0221/14/10/P10033},
year = {2019},
month = {oct},
publisher = {},
volume = {14},
number = {10},
pages = {P10033},
author = {Liu, Y. and Amjad, M.S. and Baesso, P. and Cussans, D. and Dreyling-Eschweiler, J. and Ete, R. and Gregor, I. and Huth, L. and Irles, A. and Jansen, H. and Krueger, K. and Kvasnicka, J. and Peschke, R. and Rossi, E. and Rummler, A. and Sefkow, F. and Stanitzki, M. and Wing, M. and Wu, M.},
title = {{EUDAQ2} — A flexible data acquisition software framework for common test beams},
journal = {Journal of Instrumentation}
}

@techreport{pixeltdr,
      collaboration = "ATLAS",
      author        = "ATLAS",
      title         = "{Technical Design Report for the ATLAS Inner Tracker Pixel
                       Detector}",
      institution   = "CERN",
      reportNumber  = "CERN-LHCC-2017-021, ATLAS-TDR-030",
      address       = "Geneva",
      year          = "2017",
      url           = "https://cds.cern.ch/record/2285585",
      doi           = "10.17181/CERN.FOZZ.ZP3Q",
}

@article{ibl,
    author = "Abbott, B. and others",
    collaboration = "ATLAS IBL",
    title = "{Production and Integration of the ATLAS Insertable B-Layer}",
    eprint = "1803.00844",
    archivePrefix = "arXiv",
    primaryClass = "physics.ins-det",
    reportNumber = "FERMILAB-PUB-18-826-V",
    doi = "10.1088/1748-0221/13/05/T05008",
    journal = "JINST",
    volume = "13",
    number = "05",
    pages = "T05008",
    year = "2018"
}

@article{HUGUO2010480,
title = {First reticule size {MAPS} with digital output and integrated zero suppression for the {EUDET-JRA1} beam telescope},
journal = {Nuclear Instruments and Methods in Physics Research Section A: Accelerators, Spectrometers, Detectors and Associated Equipment},
volume = {623},
number = {1},
pages = {480-482},
year = {2010},
note = {1st International Conference on Technology and Instrumentation in Particle Physics},
issn = {0168-9002},
doi = "10.1016/j.nima.2010.03.043",
url = {https://www.sciencedirect.com/science/article/pii/S0168900210006078},
author = {C. Hu-Guo and J. Baudot and G. Bertolone and A. Besson and A.S. Brogna and C. Colledani and G. Claus and R. {De Masi} and Y. Degerli and A. Dorokhov and G. Doziere and W. Dulinski and X. Fang and M. Gelin and M. Goffe and F. Guilloux and A. Himmi and K. Jaaskelainen and M. Koziel and F. Morel and F. Orsini and M. Specht and Q. Sun and O. Torheim and I. Valin and M. Winter}
}

@article{Alimonti_2025,
doi = "10.1088/1748-0221/20/03/P03024",
url = {https://doi.org/10.1088/1748-0221/20/03/P03024},
year = {2025},
month = {mar},
publisher = {IOP Publishing},
volume = {20},
number = {03},
pages = {P03024},
author = {Alimonti, G. and Andreazza, A. and Arteche, F. and Barbero, M.B. and Barrillon, P. and Beccherle, R. and Bonomelli, G. and Bilei, G.M. and Bialas, W. and Bortoletto, D. and Calderini, G. and Caratelli, A. and Cassese, A. and Christiansen, J. and Conti, E. and Crescioli, F. and Daas, M. and Damenti, L. and D'Auria, S. and De Canio, F. and De Robertis, G. and Demaria, N. and DeWitt, J. and Dieter, Y. and Dimitrievska, A. and Erdmann, W. and Esposito, S. and Exarchou, D. and Fougeron, D. and Gaioni, L. and Garcia-Sciveres, M. and Gnani, D. and Gozalez Renteria, C. and Grippo, M. and Guardino, A. and Hamer, M. and Heim, T. and Hemperek, T. and Hinterkeuser, F. and Huiberts, S. and Jara Casas, L.M. and John, J.J. and Kampkötter, J. and Karagounis, M. and Kazas, I. and Khwaira, Y. and Kluit, R. and Koukola, D. and Krieger, A. and Krüger, H. and Lalic, J. and Lauritzen, M. and Licciulli, F. and Liu, Peilian and Loddo, F. and Lopez Morillo, E. and Lounis, A. and Luongo, F. and Manghisoni, M. and Marconi, S. and Marquez Lasso, F. and Marzocca, C. and Mauer, K. and Mekkaoui, A. and Meng, Lingxin and Menichelli, M. and Menouni, M. and Minuti, M. and Mironova, M. and Miryala, S. and Missiroli, M. and Monteil, E. and Moustakas, K. and Muñoz Chavero, F. and Neue, G. and Orfanelli, S. and Paccagnella, A. and Pacher, L. and Palla, F. and Palomo Pinto, F.R. and Papadopoulou, A. and Paterno, A. and Petri, A.R. and Placidi, P. and Plackett, R. and Pradas, A. and Pulli, A. and Raciti, B. and Ratti, L. and Re, V. and Rehman, A. and Rymaszewski, P. and Sander, P. and Solal, M.C. and Standke, M. and Stugu, B. and Thompson, E. and Traversi, G. and Vogrig, D. and Vogt, M. and Wang, Tianyang and Yang, Hongtao and Zdenko, J. and The RD53 collaboration},
title = {{RD53 pixel readout integrated circuits for ATLAS and CMS HL-LHC upgrades}},
journal = {Journal of Instrumentation}
}

@article{ZurbanoFernandez:2020cco,
    author = "B{\'e}jar Alonso, I. and others",
    editor = {B{\'e}jar Alonso, I. and Br{\"u}ning, O. and Fessia, P. and Rossi, L. and Tavian, L. and Zerlauth, M.},
    title = "{High-Luminosity Large Hadron Collider (HL-LHC): Technical design report}",
    reportNumber = "CERN-2020-010",
    doi = "10.23731/CYRM-2020-0010",
    volume = "10/2020",
    month = "12",
    year = "2020",
    journal = "CERN Yellow Reports: Monographs"
}

@article{BUTTAR2025169978,
title = {{ATLAS ITk pixel detector overview}},
journal = {Nuclear Instruments and Methods in Physics Research Section A: Accelerators, Spectrometers, Detectors and Associated Equipment},
volume = {1070},
pages = {169978},
year = {2025},
issn = {0168-9002},
doi = "10.1016/j.nima.2024.169978",
url = {https://www.sciencedirect.com/science/article/pii/S0168900224009045},
author = {{C. Buttar for the ATLAS ITk Collaboration}}}

@techreport{ATL-PHYS-PUB-2021-024,
      collaboration = "ATLAS",
      title         = "{Expected tracking and related performance with the
                       updated ATLAS Inner Tracker layout at the High-Luminosity
                       LHC}",
      institution   = "CERN",
      reportNumber  = "ATL-PHYS-PUB-2021-024",
      address       = "Geneva",
      year          = "2021",
      url           = "https://cds.cern.ch/record/2776651",
      note          = "All figures including auxiliary figures are available at
                       https://atlas.web.cern.ch/Atlas/GROUPS/PHYSICS/PUBNOTES/ATL-PHYS-PUB-2021-024",
}

@misc{heim_2025_15007379,
  author       = {Heim, Timon and
                  Gallop, Bruce and
                  Alkakhi, Wael and
                  Arnaez, Olivier and
                  Crescioli, Francesco and
                  Foster, Liam and
                  Huiberts, Simon and
                  Ondrej, Kovanda and
                  Le Boulicaut, Elise and
                  Le Pottier, Luc and
                  Meng, Lingxin and
                  Mironova, Maria and
                  Pagan Griso, Simone and
                  Pianori, Elisabetta and
                  Nosler, Laura and
                  Quinn, Ryan and
                  Rastogi, Angira and
                  Rummler, Andre and
                  Tao, Zhengcheng and
                  Thompson, Emily and
                  Toldaiev, Oleksii and
                  Wittgen, Matthias},
  title        = {{YARR}},
  month        = mar,
  year         = 2025,
  publisher    = {Zenodo},
  version      = {v1.5.3},
  doi          = "10.5281/zenodo.15007379",
  url          = {https://doi.org/10.5281/zenodo.15007379},
  swhid        = {swh:1:dir:22e1f66759e0c82d560832c2028ebe4c0bfad22f
                   ;origin=https://doi.org/10.5281/zenodo.15007378;vi
                   sit=swh:1:snp:bf8fe4632f12273a06feec27fd6ca2809d23
                   7d59;anchor=swh:1:rel:cff9c678674e3e56995230f9ac38
                   a8750160ff07;path=YARR-v1.5.3
                  },
}

\end{document}